\documentclass[ reprint, onecolumn,prd,aps, 11pt, 
  ]{revtex4-1}
\usepackage{eurosym}
\usepackage{graphicx}
\usepackage{dcolumn}
\usepackage{bm}
\usepackage{amsmath,amsfonts,bm,array}
\usepackage{setspace}
\usepackage{graphicx,multirow}
\usepackage{verbatim}
\usepackage{helvet}
\usepackage{caption}
\usepackage{subcaption}

\begin{document}

\title{{\large \textbf{Exploring the quark flavor puzzle within the three-Higgs double model}}}
\author{David Emmanuel-Costa}
\email{david.costa@tecnico.ulisboa.pt}
\author{J.I. Silva-Marcos}
\email{juca@cftp.tecnico.ulisboa.pt}
\affiliation{
    Centro de F{\'\i}sica Te\'orica de Part{\'\i}culas, CFTP, 
	and Departamento de F\'{\i}sica,	
	{\it  Instituto Superior T\'ecnico, Universidade de Lisboa,}
	Avenida Rovisco Pais nr. 1, 1049-001 Lisboa, Portugal}
	
\author{Nuno Rosa Agostinho}
\email{nuno@fqa.ub.edu}

\affiliation{Departament d'Estructura i Constituents de la Mat\`eria and Institut de Ciencies del Cosmos,
Universitat de Barcelona, Diagonal 647, E-08028 Barcelona, Spain}

\begin{abstract}
We extend the standard model with two extra Higgs doublets. Making use of a
symmetry principle, we present flavour symmetries based on cycle groups $Z_N$
that oblige each Higgs doublet to contribute to the mass of only one
generation. The Higgs doublets couple to the fermions with different
strengths and in this way accommodate the quark mass hierarchy. We
systematically search for all charge configurations that naturally lead to
the alignment in flavour space of the quark sectors, resulting in a CKM
matrix near to the identity, determined by the quark mass hierarchy, and
with the correct overall phenomenological features. The minimal realization
is by the group $Z_7$. We show that only a limited number of solutions
exist, and that any accidental global symmetry that may occur together with
the discrete symmetry is necessarily anomalous. A phenomenological study of
each class of solutions concerning predictions to the flavour changing
neutral current (FCNC) phenomena is also performed: for some solutions, it
is possible to obtain realistic quark masses and mixing, while the flavour
violating neutral Higgs are light enough to be accessible at LHC.
\end{abstract}

\maketitle


%\tableofcontents

\section{Introduction}

The discovery of the Higgs boson in 2012 at the LHC has attested the success
of the standard model (SM) in describing the observed fermions and their
interactions. However, there exist many theoretical issues or open questions
that have no satisfactory answer. In particular, the observed flavour
pattern lacks of a definitive explanation, i.e., the quark Yukawa coupling
matrices $Y_u$ and $Y_d$, which in the SM reproduce the six quark masses,
three mixings angles and a complex phase to account for CP violation
phenomena, are general complex matrices, not constrained by any gauge
symmetry.

Experimentally the flavour puzzle is very intricate. First, there is the
quark mass hierarchy in both sectors. Secondly, the mixings in the SM,
encoded in the Cabibbo-Kobayashi-Maskawa (CKM) unitary matrix, turns out to
be close to the identity matrix. If one takes also the lepton sector into
account, the hierarchy there is even more puzzling~\cite%
{Emmanuel-Costa:2015tca}. On the other hand, in the SM there is in general
no connection between the quark masses hierarchy and the CKM mixing pattern.
In fact, if one considers the Extreme Chiral Limit, where the quark masses
of the first two generations are set to zero, the mixing does not
necessarily vanish~\cite{Botella:2016krk}, and one concludes that the CKM
matrix~$V$ being close to the identity matrix has nothing to do with the
fact that the quark masses are hierarchical. Indeed, in order to have $%
V\approx \mathbf{1}$, one must have a definite alignment of the quark mass
matrices in the flavour space, and to explain this alignment, a flavour
symmetry or some other mechanism is required~\cite{Botella:2016krk}.

Among many attempts made in the literature to address the flavour puzzle,
extensions of the SM with new Higgs doublet are particularly motivating.
This is due to fact that the number of Higgs doublets is not constrained by
the SM symmetry. Moreover, the addition of scalar doublets gives rise to new
Yukawa interactions and as a result it provides a richer framework in
approaching the theory of flavour. On the other hand, any new extension of
the Higgs sector must be very much constrained, since it naturally leads to
flavour changing neutral currents. At tree level, in the SM, all the flavour
changing transitions are mediated through charged weak currents and the
flavour mixing is controlled by the CKM matrix~\cite%
{Cabibbo:1963yz,Kobayashi:1973fv}. If new Higgs doublets are added, one
expects large FCNC effects already present at tree level. Such effects have
not been experimentally observed and they constrain severely any model with
extra Higgs doublets, unless a flavour symmetry suppresses or avoids large
FCNC~\cite{Branco:2011iw}.

Minimal flavour violating models~\cite%
{Joshipura:1990pi,Antaramian:1992ya,Hall:1993ca,Mantilla2017, Buras:2001,
Dambrosio:2002} are examples of a multiHiggs extension where FCNC are
present at tree-level but their contributions to FCNC phenomena involve only
off-diagonal elements of the CKM matrix or their products. The first
consistent models of this kind were proposed by Branco, Grimus and Lavoura
(BGL)~\cite{Branco:1996bq}, and consisted of the SM with two Higgs doublets
together with the requirement of an additional discrete symmetry. BGL models
are compatible with lower neutral Higgs masses and FCNC's occur at tree
level, with the new interactions entirely determined in terms of the CKM
matrix elements.

The goal of this paper is to generalize the previous BGL models and to,
systematically, search for patterns where a discrete flavour symmetry
naturally leads to the alignment of the flavour space of both the quark
sectors. Although the quark mass hierarchy does not arise from the symmetry,
the effect of both is such that the CKM matrix is near to the identity and
has the correct overall phenomenological features, determined by the quark
mass hierarchy, \cite{Branco:2011aa}. To do this we extend the SM with two
extra Higgs doublets to a total of three Higgs $\phi _{a}$. The choice for
discrete symmetries is to avoid the presence of Goldstone bosons that appear
in the context of any global continuous symmetry, when the spontaneous
electroweak symmetry breaking occurs. For the sake of simplicity, we
restrict our search to the family group $Z_{N}$, and demand that the
resulting up-quark mass matrix $M_{u}$ is diagonal. This is to say that, due
to the expected strong up-quark mass hierarchy, we only consider those cases
where the contribution of the up-quark mass matrix to quark mixing is
negligible.

If one assumes that all Higgs doublets acquire vacuum expectation values
with the same order of magnitude, then each Higgs doublet must couple to the
fermions with different strengths. Possibly one could obtain similar results
assuming that the vacuum expectation values (VEVs) of the Higgs have a
definite hierarchy instead of the couplings, but this is not considered
here. Combining this assumption with the symmetry, we obtain the correct
ordered hierarchical pattern if the coupling with $\phi _{3}$ gives the
strength of the third generation, the coupling with $\phi _{2}$ gives the
strength of the second generation and the coupling with $\phi _{1}$ gives
the strength of the first generation. Therefore, from our point of view, the
three Higgs doublets are necessary to ensure that there exists three
different coupling strengths, one for each generation, to guarantee
simultaneously an hierarchical mass spectrum and a CKM matrix that has the
correct overall phenomenological features e.g. $\left\vert V_{cb}\right\vert
^{2}+\left\vert V_{ub}\right\vert ^{2}=O(m_{s}/m_{b})^{2}$, \ and denoted
here by $V\approx \mathbf{1.}$

Indeed, our approach is within the BGL models, and such that the FCNC
flavour structure is entirely determined by CKM. Through the symmetry, the
suppression of the most dangerous FCNC's, by combinations of the CKM matrix
elements and light quark masses, is entirely natural.

The paper is organised as follows. In the next section, we present our model
and classify the patterns allowed by the discrete symmetry in combination with our assumptions. 
In Sec. \ref{sec:num}, we give a brief numerical analysis of
the phenomenological output of our solutions. In Sec. \ref{sec:fcnc}, we
examine the suppression of scalar mediated FCNC in our framework for each
pattern. Finally, in Sec. \ref{sec:conc}, we present our conclusions.

\section{The Model}

\label{sec:model}

We extend the Higgs sector of the SM with two extra new scalar doublets,
yielding a total of three scalar doublets, as $\phi _{1}$, $\phi _{2}$, $%
\phi _{3}$. As it was mentioned in the introduction, the main idea for
having three Higgs doublets is to implement a discrete flavour symmetry,
that leads to the alignment of the flavour space of the quark sectors. The
quark mass hierarchy does not arise from the symmetry, but together with the
symmetry the effect of both is such that the CKM matrix is near to the
identity and has the correct overall phenomenological features, determined
by the quark mass hierarchy.

Let us start by considering the most general quark Yukawa coupling
Lagrangian invariant in our setup 
\begin{equation}
-\mathcal{L}_{\text{Y}}=(\Omega _{a})_{ij}\,\overline{Q}_{Li}\ \widetilde{%
\phi }_{a}\ u_{R_{j}}+(\Gamma _{a})_{ij}\,\overline{Q}_{Li}\ \phi _{a}\
d_{R_{j}}+h.c.,  \label{eq:lag}
\end{equation}%
with the Higgs labeling $a=1,2,3$ and $i,j$ are just the usual flavour
indexes identifying the generations of fermions. In the above Lagrangian,
one has three Yukawa coupling matrices $\Omega _{1}$, $\Omega _{2}$, $\Omega
_{3}$ for the up-quark sector and three Yukawa coupling matrices $\Gamma
_{1} $, $\Gamma _{2}$, $\Gamma _{3}$ for the down sector, corresponding to
each of the Higgs doublets $\phi _{1}$, $\phi _{2}$, $\phi _{3}$. Assuming
that only the neutral components of the three Higgs doublets acquire vacuum
expectation value (VEV), the quark mass $M_{u}$ and $M_{d}$ are then easily
generated as 
\begin{subequations}
\label{eq:mass}
\begin{align}
M_{u}& =\Omega _{1}\left\langle \phi _{1}\right\rangle \,^{\ast }+\,\Omega
_{2}\left\langle \phi _{2}\right\rangle \,^{\ast }+\,\Omega
_{3}\,\left\langle \phi _{3}\right\rangle ^{\ast },  \label{eq:massup} \\
M_{d}& =\Gamma _{1}\left\langle \phi _{1}\right\rangle \,+\,\Gamma
_{2}\left\langle \phi _{2}\right\rangle \,+\,\Gamma _{3}\left\langle \phi
_{3}\right\rangle ,
\end{align}%
where VEVs $\langle \phi _{i}\rangle $ are parametrised as 
\end{subequations}
\begin{equation}
\left\langle \phi _{1}\right\rangle =\frac{v_{1}}{\sqrt{2}},\quad
\left\langle \phi _{2}\right\rangle =\frac{v_{2}e^{i\alpha _{2}}}{\sqrt{2}}%
,\quad \left\langle \phi _{3}\right\rangle =\frac{v_{3}e^{i\alpha _{3}}}{%
\sqrt{2}},
\end{equation}%
with $v_{1}$, $v_{2}$ and $v_{3}$ being the VEV moduli and $\alpha _{2}$, $%
\alpha _{3}$ just complex phases. We have chosen the VEV of $\phi _{1}$ to
be real and positive, since this is always possible through a proper gauge
transformation. As stated, we assume that the moduli of VEVs $v_{i}$ are of
the same order of magnitude, i.e., 
\begin{equation}
v_{1}\sim v_{2}\sim v_{3}.  \label{vs}
\end{equation}

Each of the $\phi _{a}$ couples to the quarks with a coupling $(\Omega
_{a})_{ij},(\Gamma _{a})_{ij}$ which we take be of the same order of
magnitude, unless some element vanishes by imposition of the flavour
symmmetry. In this sence, each $\phi _{a}$ and $(\Omega _{a},\Gamma _{a})$
will generate it's own respective generation: i.e., our model is such that
by imposition of the flavour symmmetry, $\phi _{3}$, $\Omega _{3}$, $\Gamma
_{3}$ will generate $m_{t}$ respectivelly $m_{b}$, that $\phi _{2}$, $\Omega
_{2}$, $\Gamma _{2}$ will generate $m_{c}$ respectivelly $m_{s}$, and that $%
\phi _{1}$, $\Omega _{1}$, $\Gamma _{1}$ will generate $m_{u}$ respectivelly 
$m_{d}$. Generically, we have 
\begin{subequations}
\label{eq:hierarchy}
\begin{align}
v_{1}\left\vert (\Omega _{1})_{ij}\right\vert & \sim m_{u},\;v_{2}\left\vert
(\Omega _{2})_{ij}\right\vert \sim m_{c},\;v_{3}\left\vert (\Omega
_{3})_{ij}\right\vert \sim m_{t}, \\
v_{1}\left\vert (\Gamma _{1})_{ij}\right\vert & \sim m_{d},\;v_{2}\left\vert
(\Gamma _{2})_{ij}\right\vert \sim m_{s},\;v_{3}\left\vert (\Gamma
_{3})_{ij}\right\vert \sim m_{b},
\end{align}%
which together with Eq. \ref{vs} implies a definite hierarchy amongst the
non-vanishing Yukawa coupling matrix elements: 
\end{subequations}
\begin{subequations}
\label{eq:hier}
\begin{align}
\left\vert (\Omega _{1})_{ij}\right\vert & \ll \left\vert (\Omega
_{2})_{ij}\right\vert \ll \left\vert (\Omega _{3})_{ij}\right\vert ,
\label{eq:hierup} \\[2mm]
\left\vert (\Gamma _{1})_{ij}\right\vert & <\left\vert (\Gamma
_{2})_{ij}\right\vert \ll \left\vert (\Gamma _{3})_{ij}\right\vert .
\end{align}

Next, we focus on the required textures for the Yukawa coupling matrices $%
\Omega _{a}$ and $\Gamma _{a}$ that naturally lead to an hierarchical mass
quark spectrum and at the same time to a realistic CKM mixing matrix. These
textures must be reproduced by our choice of the flavour symmetry. As
referred in the introduction, we search for quark mass patterns where the
mass matrix $M_{u}$ is diagonal. Therefore, one derives from Eqs.~%
\eqref{eq:massup}, \eqref{eq:hierup} the following textures for $\Omega _{a}$
\end{subequations}
\begin{equation}
\Omega _{1}=%
\begin{pmatrix}
\mathsf{x} & 0 & 0 \\ 
0 & 0 & 0 \\ 
0 & 0 & 0%
\end{pmatrix}%
,\,\Omega _{2}=%
\begin{pmatrix}
0 & 0 & 0 \\ 
0 & \mathsf{x} & 0 \\ 
0 & 0 & 0%
\end{pmatrix}%
,\,\Omega _{3}=%
\begin{pmatrix}
0 & 0 & 0 \\ 
0 & 0 & 0 \\ 
0 & 0 & \mathsf{x}%
\end{pmatrix}%
.  \label{eq:textureOs}
\end{equation}%
The entry $\mathsf{x}$ means a non zero element. In this case, the up-quark
masses are given by $m_{u}=v_{1}\left\vert (\Omega _{1})_{11}\right\vert $, $%
m_{c}=v_{2}\left\vert (\Omega _{2})_{22}\right\vert $ and $%
m_{t}=v_{3}\left\vert (\Omega _{3})_{33}\right\vert $.

Generically, the down-quark Yukawa coupling matrices must have the following
indicative textures 
\begin{equation}
\Gamma _{1}=%
\begin{pmatrix}
\boldsymbol{\mathsf{x}} & \boldsymbol{\mathsf{x}} & \boldsymbol{\mathsf{x}}
\\ 
\mathsf{x} & \mathsf{x} & \mathsf{x} \\ 
\mathsf{x} & \mathsf{x} & \mathsf{x}%
\end{pmatrix}%
,\,\Gamma _{2}=%
\begin{pmatrix}
0 & 0 & 0 \\ 
\boldsymbol{\mathsf{x}} & \boldsymbol{\mathsf{x}} & \boldsymbol{\mathsf{x}}
\\ 
\mathsf{x} & \mathsf{x} & \mathsf{x}%
\end{pmatrix}%
,\,\Gamma _{3}=%
\begin{pmatrix}
0 & 0 & 0 \\ 
0 & 0 & 0 \\ 
\boldsymbol{\mathsf{x}} & \boldsymbol{\mathsf{x}} & \boldsymbol{\mathsf{x}}%
\end{pmatrix}%
.  \label{eq:textureGs}
\end{equation}%
We distinguish rows with bold $\boldsymbol{\mathsf{x}}$ in order to indicate
that it is mandatory that at least one of matrix elements within that row
must be nonvanishing. Rows denoted with $\mathsf{x}$ may be set to zero,
without modifying the mass matrix hierarchy. These textures ensure that not only is 
the mass spectrum hierarchy respected but it also leads to the alignment of the 
flavor space of both the quark sectors \cite{Branco:2011aa}
and to a CKM matrix $V\approx \mathbf{1}$. For instance, if one would not have
a vanishing, or comparatively very small, $(1,3)$ entry in the $\Gamma _{2}$%
, this would not necessarily spoil the scale of $m_{s}$, but it would
dramatically change the predictions for the CKM mixing matrix.

In order to force the Yukawa coupling matrices $\Omega _{a}$ and $\Gamma
_{a} $ to have the indicative forms outlined in Eqs.~\eqref{eq:textureOs}
and~\eqref{eq:textureGs}, we introduce a global flavour symmetry. Since any
global continuous symmetry leads to the presence of massless Goldstone
bosons after the spontaneous electroweak breaking, one should instead
consider a discrete symmetry. Among many possible discrete symmetry
constructions, we restrict our searches to the case of cycle groups $Z_{N}$.
Thus, we demand that any quark or boson multiplet $\chi $ transforms
according to $Z_{N}$ as 
\begin{equation}
\chi \rightarrow \chi ^{\prime }=e^{i\,\mathcal{Q}(\chi )\,\frac{2\pi }{N}%
}\chi ,
\end{equation}%
where $\mathcal{Q}(\chi )\in \{0,1,\dots ,N\}$ is the $Z_{N}$-charge
attributed for the multiplet $\chi $.

We have chosen the up-quark mass matrix $M_{u}$ to be diagonal. This
restricts the flavour symmetry $Z_{N}$. We have found that, in order to
ensure that all Higgs doublet charges are different, and to
have appropriate charges for fields ${Q_{L}}_{i}$ and ${u_{R}}_{i}$, we must
have $N\geq 7$. We simplify our analysis by fixing $N=7$ and choose: 
\begin{subequations}
\label{eq:fix}
\begin{align}
\mathcal{Q}({Q_{L}}_{i})& =(0,1,-2), \\
\mathcal{Q}({u_{R}}_{i})& =(0,2,-4),
\end{align}%
In addition, we may also fix 
\end{subequations}
\begin{equation}
\mathcal{Q}({Q_{L}}_{i})=\mathcal{Q}(\phi _{i})  \label{eq:fix1}
\end{equation}%
It turns out that these choices do not restrict the results, i.e. the
possible textures that one can have for the $\Gamma _{i}$ matrices. Other
choices would only imply that we reshuffle the charges of the multiplets.

With the purpose of enumerating the different possible textures for the $%
\Gamma _{i}$ matrices implementable in $Z_{7}$, we write down the charges of
the trilinears $\mathcal{Q}({\overline{Q}_{L}}_{i}\phi _{a}{d_{R}}_{j})$
corresponding to each $\phi _{a}$ as 
\begin{subequations}
\begin{equation}
\mathcal{Q}({\overline{Q}_{L}}_{i}\phi _{1}{d_{R}}_{j})=%
\begin{pmatrix}
d_{1} & d_{2} & d_{3} \\ 
d_{1}-1 & d_{2}-1 & d_{3}-1 \\ 
d_{1}+2 & d_{2}+2 & d_{3}+2%
\end{pmatrix}%
,
\end{equation}%
\begin{equation}
\mathcal{Q}({\overline{Q}_{L}}_{i}\phi _{2}{d_{R}}_{j})=%
\begin{pmatrix}
d_{1}+1 & d_{2}+1 & d_{3}+1 \\ 
d_{1} & d_{2} & d_{3} \\ 
d_{1}+3 & d_{2}+3 & d_{3}+3%
\end{pmatrix}%
,
\end{equation}%
\begin{equation}
\mathcal{Q}({\overline{Q}_{L}}_{i}\phi _{3}{d_{R}}_{j})=%
\begin{pmatrix}
d_{1}-2 & d_{2}-2 & d_{3}-2 \\ 
d_{1}-3 & d_{2}-3 & d_{3}-3 \\ 
d_{1} & d_{2} & d_{3}%
\end{pmatrix}%
,
\end{equation}%
where $d_{i}\equiv \mathcal{Q}({d_{R}}_{i})$. One can check that, in order
to have viable solutions, one must vary the values of $d_{i}\in
\{0,1,-2,-3\} $.

We summarise in Table \ref{tab:downTextures} all the allowed textures for
the $\Gamma _{a}$ matrices and the resulting $M_{d}$ mass matrix texture,
excluding all cases which are irrelevant, e.g. matrices that have too much
texture zeros and are singular, or matrices that do not accommodate CP
violation. It must be stressed, that these are the textures obtained by the
different charge configurations that one can possibly choose. However, if
one assumes a definite charge configuration, then the entire texture, $M_{d}$
and $M_{u}$ and the respective phenomenology are fixed. As stated, the list
of textures in Table~\ref{tab:downTextures} remains unchanged even if one
chooses any other set than in Eqs.~\eqref{eq:fix}, \eqref{eq:fix1}. As
stated, that all patterns presented here are of the Minimal Flavour
Violation (MFV) type \cite%
{Joshipura:1990pi,Antaramian:1992ya,Hall:1993ca,Mantilla2017, Buras:2001,
Dambrosio:2002}.

Pattern~I in the table was already considered in Ref.~\cite{Botella:2009pq}
in the context of $Z_{8}$. We discard Patterns~IV, VII and X, because
contrary to our starting point, at least one of three non-zero couplings
with $\phi _{1}$ will turn out be of the same order as the larger coupling
with $\phi _{2}$ in order to meet the phenomenological requirements of the
CKM matrix.

Notice also, that the structure of other $M_{d}$'s cannot be trivially
obtained, e.g. from Pattern I, by a transformation of the right-handed down
quark fields.

Our symmetry model may be extended to the charged leptons and neutrinos,
e.g. in the context of type one see-saw. Choosing for the lepton doublets ${L%
}_{i}$ the charges $\mathcal{Q}({L}_{i})=(0,-1,2)$, opposite to the Higgs
doublets in Eq. \eqref{eq:fix1},~and e.g. for the charges $\mathcal{Q}({e_{R}%
}_{i})=(0,-2,4)$ of the right-handed fields ${e_{R}}_{i}$, we force the
charged~ lepton mass matrix to be diagonal. Then for the right-handed
neutrinos ${\nu _{R}}_{i}$, choosing $\mathcal{Q}({\nu _{R}}_{i})=(0,0,0)$,
we obtain for the neutrino Dirac mass matrix a pattern similar to pattern I.
Of course, for this case, the heavy right-handed neutrino Majorana mass
matrix is totally arbitrary. In other cases, i.e. for other patterns and
charges, in particular for the right-handed neutrinos, we could introduce
scalar singlets with suitable charges, which would then lead to certain
heavy right-handed neutrino Majorana mass matrices.

Next, we address an important issue of the model, namely, whether accidental 
$U(1)$ symmetries may appear in the Yukawa sector or in the potential. One
may wonder whether a continuous accidental $U(1)$ symmetry could arise, once
the $Z_{7}$ is imposed at the Lagrangian level in Eq.~\eqref{eq:lag}. This
is indeed the case, i.e., for all realizations of $Z_{7}$, one has the
appearance of a global $U(1)_{X}$. However, any consistent global $U(1)_{X}$
must obey to the anomaly-free conditions of global symmetries~\cite%
{Babu:1989ex}, which read for the anomalies $SU(3)^{2}\times U(1)_{X}$, $%
SU(2)^{2}\times U(1)_{X}$ and $U(1)_{Y}^{2}\times U(1)_{X}$ as 
\end{subequations}
\begin{subequations}
\begin{equation}
A_{3}\equiv \frac{1}{2}\sum_{i=1}^{3}\biggl(2X({Q_{L}}_{i})-X({u_{R}}_{i})-X(%
{d_{R}}_{i})\biggr)=0,  \label{eq:A3}
\end{equation}%
\begin{equation}
A_{2}\equiv \frac{1}{2}\sum_{i=1}^{3}\biggl(3X({Q_{L}}_{i})+X({\ell _{L}}%
_{i})\biggr)=0,  \label{eq:A2}
\end{equation}%
\begin{equation}
A_{1}\equiv \frac{1}{6}\sum_{i=1}^{3}\biggl(X({Q_{L}}_{i})+3X({\ell _{L}}%
_{i})-8X({u_{R}}_{i})-2X({d_{R}}_{i})-6X({e_{R}}_{i})\biggr)=0,
\end{equation}%
where $X(\chi )$ is the $U(1)_{X}$ charge of the fermion multiplet $\chi $.
We have properly shifted the $Z_{7}$-charges in Eq.\eqref{eq:fix} and in
Table \ref{tab:downTextures} so that $X(\chi )=\mathcal{Q}(\chi )$, apart of
an overall $U(1)_{X}$ convention. In general to test those conditions, one
needs to specify the transformation laws for all fermionic fields. Looking
at the Table~1, we derive that all the cases, except the first case
corresponding to $d_{i}=(0,0,0)$, violate the condition given in Eq.~%
\eqref{eq:A3} that depends only on coloured fermion multiplets. In the case $%
d_{i}=(0,0,0)$, if one assigns the charged lepton charges as $X({\ell _{L}}%
_{i})=X({Q_{L}}_{i})$ one concludes that the condition given in Eq.~%
\eqref{eq:A2} is violated. One then concludes that the global $U(1)_{X}$
symmetry is anomalous and therefore only the discrete symmetry $Z_{7}$
persists.

We also comment on the scalar potential of our model. The most general
scalar potential with three scalars invariant under $Z_{7}$ reads as 
\end{subequations}
\begin{equation}
V(\phi )=\sum_{i}\left[ -\mu _{i}^{2}\phi _{i}^{\dagger }\phi _{i}+\lambda
_{i}(\phi _{i}^{\dagger }\phi _{i})^{2}\right] +\sum_{i<j}\left[ +C_{i}(\phi
_{i}^{\dagger }\phi _{i})(\phi _{j}^{\dagger }\phi _{j})+\,\bar{C}%
_{i}\left\vert \phi _{i}^{\dagger }\phi _{j}\right\vert ^{2}\right] ,
\label{eq:pot}
\end{equation}%
where the constants $\mu _{i}^{2}$, $\lambda _{i}$, $C_{i}$ and $\bar{C}_{i}$
are taken real for $i,j=1,2,3$. Analysing the potential above, one sees that
it gives rise to the accidental global continuous symmetry $\phi
_{i}\rightarrow e^{i\alpha _{i}}\phi _{i}$, for arbitrary $\alpha _{i}$,
which upon spontaneous symmetry breaking leads to a massless neutral scalar,
at tree level. Introducing soft-breaking terms like $m_{ij}^{2}\phi
_{i}^{\dagger }\,\phi _{j}\,+\text{H.c.}$ can erase the problem. Another
possibility without spoiling the $Z_{7}$ symmetry is to add new scalar
singlets, so that the coefficients $m_{ij}^{2}$ are effectively obtained
once the scalar singlets acquire VEVs.

%%%%%%%%%%%%%%%%%%%%%%%%%%%%%%%%%%%%%%%%%%%%%%%%%%%%%%%%%%%%%%%%%%%%%%

\begin{table}[]
\caption{The table shows the viable configurations for the right-handed
down-quark ${d_{R}}_{i}$ and their corresponding $\Gamma _{1}$, $\Gamma _{2}$%
, $\Gamma_{3}$ and $M_{d}$ matrices. It is understood that, for each pattern
and coupling, the parameters expressed here by the same symbol, are in fact
different, but denoting he same order of magnitude, (or possibly smaller).
E.g. in pattern I, coupling $\Gamma_1$, the three $\protect\delta$, $\protect%
\delta$, $\protect\delta$, stand for $\protect\delta_1$, $\protect\delta_2$, 
$\protect\delta_3$. The same applies to the $\protect\varepsilon$'s and $c$%
's. For patterns IV, VII, and X, which will be excluded, one of the
couplings in $\Gamma _{1}$ turns out to be much larger. }
\label{tab:downTextures}\centering
\begin{tabular}{|c|c|c|c|c|c|}
\hline
Pattern & $\mathcal{Q}({d_R}_i)$ & $\Gamma_1$ & $\Gamma_2$ & $\Gamma_3$ & $%
M_d$ \\ \hline
I & $(0, 0, 0)$ & $%
\begin{pmatrix}
\delta & \delta & \delta \\ 
0 & 0 & 0 \\ 
0 & 0 & 0%
\end{pmatrix}%
$ & $%
\begin{pmatrix}
0 & 0 & 0 \\ 
\varepsilon & \varepsilon & \varepsilon \\ 
0 & 0 & 0%
\end{pmatrix}%
$ & $%
\begin{pmatrix}
0 & 0 & 0 \\ 
0 & 0 & 0 \\ 
c & c & c%
\end{pmatrix}%
$ & $%
\begin{pmatrix}
\delta & \delta & \delta \\ 
\varepsilon & \varepsilon & \varepsilon \\ 
c & c & c%
\end{pmatrix}%
$ \rule{0pt}{0.9cm}\rule[-0.9cm]{0pt}{0pt} \\ 
II & $(0, 0, 1)$ & $%
\begin{pmatrix}
\delta & \delta & 0 \\ 
0 & 0 & \delta \\ 
0 & 0 & 0%
\end{pmatrix}%
$ & $%
\begin{pmatrix}
0 & 0 & 0 \\ 
\varepsilon & \varepsilon & 0 \\ 
0 & 0 & 0%
\end{pmatrix}%
$ & $%
\begin{pmatrix}
0 & 0 & 0 \\ 
0 & 0 & 0 \\ 
c & c & 0%
\end{pmatrix}%
$ & $%
\begin{pmatrix}
\delta & \delta & 0 \\ 
\varepsilon & \varepsilon & \delta \\ 
c & c & 0%
\end{pmatrix}%
$ \rule{0pt}{0.9cm}\rule[-0.9cm]{0pt}{0pt} \\ 
III & $(0, 0, -3)$ & $%
\begin{pmatrix}
\delta & \delta & 0 \\ 
0 & 0 & 0 \\ 
0 & 0 & 0%
\end{pmatrix}%
$ & $%
\begin{pmatrix}
0 & 0 & 0 \\ 
\varepsilon & \varepsilon & 0 \\ 
0 & 0 & \varepsilon%
\end{pmatrix}%
$ & $%
\begin{pmatrix}
0 & 0 & 0 \\ 
0 & 0 & 0 \\ 
c & c & 0%
\end{pmatrix}%
$ & $%
\begin{pmatrix}
\delta & \delta & 0 \\ 
\varepsilon & \varepsilon & 0 \\ 
c & c & \varepsilon%
\end{pmatrix}%
$ \rule{0pt}{0.9cm}\rule[-0.9cm]{0pt}{0pt} \\ 
IV & $(0, 0, -2)$ & $%
\begin{pmatrix}
\delta & \delta & 0 \\ 
0 & 0 & 0 \\ 
0 & 0 & \varepsilon%
\end{pmatrix}%
$ & $%
\begin{pmatrix}
0 & 0 & 0 \\ 
\varepsilon & \varepsilon & 0 \\ 
0 & 0 & 0%
\end{pmatrix}%
$ & $%
\begin{pmatrix}
0 & 0 & 0 \\ 
0 & 0 & 0 \\ 
c & c & 0%
\end{pmatrix}%
$ & $%
\begin{pmatrix}
\delta & \delta & 0 \\ 
\varepsilon & \varepsilon & 0 \\ 
c & c & \varepsilon%
\end{pmatrix}%
$ \rule{0pt}{0.9cm}\rule[-0.9cm]{0pt}{0pt} \\ 
V & $(0, 1, 0)$ & $%
\begin{pmatrix}
\delta & 0 & \delta \\ 
0 & \delta & 0 \\ 
0 & 0 & 0%
\end{pmatrix}%
$ & $%
\begin{pmatrix}
0 & 0 & 0 \\ 
\varepsilon & 0 & \varepsilon \\ 
0 & 0 & 0%
\end{pmatrix}%
$ & $%
\begin{pmatrix}
0 & 0 & 0 \\ 
0 & 0 & 0 \\ 
c & 0 & c%
\end{pmatrix}%
$ & $%
\begin{pmatrix}
\delta & 0 & \delta \\ 
\varepsilon & \delta & \varepsilon \\ 
c & 0 & c%
\end{pmatrix}%
$ \rule{0pt}{0.9cm}\rule[-0.9cm]{0pt}{0pt} \\ 
VI & $(0, -3, 0)$ & $%
\begin{pmatrix}
\delta & 0 & \delta \\ 
0 & 0 & 0 \\ 
0 & 0 & 0%
\end{pmatrix}%
$ & $%
\begin{pmatrix}
0 & 0 & 0 \\ 
\varepsilon & 0 & \varepsilon \\ 
0 & \varepsilon & 0%
\end{pmatrix}%
$ & $%
\begin{pmatrix}
0 & 0 & 0 \\ 
0 & 0 & 0 \\ 
c & 0 & c%
\end{pmatrix}%
$ & $%
\begin{pmatrix}
\delta & 0 & \delta \\ 
\varepsilon & 0 & \varepsilon \\ 
c & \varepsilon & c%
\end{pmatrix}%
$ \rule{0pt}{0.9cm}\rule[-0.9cm]{0pt}{0pt} \\ 
VII & $(0, -2, 0)$ & $%
\begin{pmatrix}
\delta & 0 & \delta \\ 
0 & 0 & 0 \\ 
0 & \varepsilon & 0%
\end{pmatrix}%
$ & $%
\begin{pmatrix}
0 & 0 & 0 \\ 
\varepsilon & 0 & \varepsilon \\ 
0 & 0 & 0%
\end{pmatrix}%
$ & $%
\begin{pmatrix}
0 & 0 & 0 \\ 
0 & 0 & 0 \\ 
c & 0 & c%
\end{pmatrix}%
$ & $%
\begin{pmatrix}
\delta & 0 & \delta \\ 
\varepsilon & 0 & \varepsilon \\ 
c & \varepsilon & c%
\end{pmatrix}%
$ \rule{0pt}{0.9cm}\rule[-0.9cm]{0pt}{0pt} \\ 
VIII & $(1, 0, 0)$ & $%
\begin{pmatrix}
0 & \delta & \delta \\ 
\delta & 0 & 0 \\ 
0 & 0 & 0%
\end{pmatrix}%
$ & $%
\begin{pmatrix}
0 & 0 & 0 \\ 
0 & \varepsilon & \varepsilon \\ 
0 & 0 & 0%
\end{pmatrix}%
$ & $%
\begin{pmatrix}
0 & 0 & 0 \\ 
0 & 0 & 0 \\ 
0 & c & c%
\end{pmatrix}%
$ & $%
\begin{pmatrix}
0 & \delta & \delta \\ 
\delta & \varepsilon & \varepsilon \\ 
0 & c & c%
\end{pmatrix}%
$ \rule{0pt}{0.9cm}\rule[-0.9cm]{0pt}{0pt} \\ 
IX & $(-3, 0, 0)$ & $%
\begin{pmatrix}
0 & \delta & \delta \\ 
0 & 0 & 0 \\ 
0 & 0 & 0%
\end{pmatrix}%
$ & $%
\begin{pmatrix}
0 & 0 & 0 \\ 
0 & \varepsilon & \varepsilon \\ 
\varepsilon & 0 & 0%
\end{pmatrix}%
$ & $%
\begin{pmatrix}
0 & 0 & 0 \\ 
0 & 0 & 0 \\ 
0 & c & c%
\end{pmatrix}%
$ & $%
\begin{pmatrix}
0 & \delta & \delta \\ 
0 & \varepsilon & \varepsilon \\ 
\varepsilon & c & c%
\end{pmatrix}%
$ \rule{0pt}{0.9cm}\rule[-0.9cm]{0pt}{0pt} \\ 
X & $(-2, 0, 0)$ & $%
\begin{pmatrix}
0 & \delta & \delta \\ 
0 & 0 & 0 \\ 
\varepsilon & 0 & 0%
\end{pmatrix}%
$ & $%
\begin{pmatrix}
0 & 0 & 0 \\ 
0 & \varepsilon & \varepsilon \\ 
0 & 0 & 0%
\end{pmatrix}%
$ & $%
\begin{pmatrix}
0 & 0 & 0 \\ 
0 & 0 & 0 \\ 
0 & c & c%
\end{pmatrix}%
$ & $%
\begin{pmatrix}
0 & \delta & \delta \\ 
0 & \varepsilon & \varepsilon \\ 
\varepsilon & c & c%
\end{pmatrix}%
$\rule{0pt}{0.9cm}\rule[-0.9cm]{0pt}{0pt} \\ \hline
\end{tabular}%
\end{table}
\newpage

\section{Numerical analysis}

\label{sec:num}

In this section, we give the phenomenological predictions obtained by the
patterns listed in Table~\ref{tab:downTextures}. Note that, although these
patterns arrize directly from the chosen discrete charge configuration of
the quark fields, one may further preform a residual flavour transformation
of the right-handed down quark fields, resulting in an extra zero entry in $%
M_{d}$. Taking this into account, all the parameters in each pattern may be
uniquely expressed in terms of down quark masses and the CKM matrix elements 
$V_{ij}$. This follows directly from the diagonalization equation of $M_{d}:$
\begin{equation}
V\ ^{\dagger }M_{d}\ W=diag(m_{d},m_{s},m_{b})\quad \Longrightarrow \quad
M_{d}=V\ diag(m_{d},m_{s},m_{b})\ W^{\dagger }  \label{diag}
\end{equation}%
with $V$ being the CKM mixing matrix, since $M_{u}$ is diagonal. Because of
the zero entries in $M_{d}$, it is easy to extract the right-handed
diagonalization matrix $W$, completely in terms of the down quark masses and
the $V_{ij}$. Thus, all parameters, modulo the residual transformation of
the right-handed down quark fields, are fixed, i.e., all parameters in each
pattern may be uniquely expressed in terms of down quark masses and the CKM
matrix elements $V_{ij}$, including the right-handed diagonalization matrix $%
W$ of $M_{d}$. More precisely, all matrix elements of $V$ are written in
terms of Wolfenstein real parameters $\lambda $, $A$, $\overline{\rho }$ and 
$\overline{\eta }$, defined in terms of rephasing invariant quantities as 
\begin{subequations}
\begin{equation}
\lambda \equiv \frac{|V_{us}|}{\sqrt{|V_{us}|^{2}+|V_{ud}|^{2}}},\qquad
A\equiv \frac{1}{\lambda }\left\vert \frac{V_{cb}}{V_{us}}\right\vert \,,
\end{equation}%
\begin{equation}
\overline{\rho }+i\,\overline{\eta }\equiv -\frac{V_{ud}^{\phantom{\ast}%
}V_{ub}^{\ast }}{V_{cd}^{\phantom{\ast}}V_{cb}^{\ast }}
\end{equation}%
and $diag(m_{d},m_{s},m_{b})$ in Eq.~\eqref{diag} 
\end{subequations}
\begin{equation}
\begin{array}{l}
\sqrt{\frac{m_{d}}{m_{s}}}=\sqrt{{\frac{k_{d}}{k_s}}}\ \lambda \\ 
\\ 
\frac{m_{s}}{m_{b}}=k_{s}\ \lambda ^{2}%
\end{array}%
\quad \Longrightarrow \quad 
\begin{array}{l}
m_{d}=k_{d}\ \lambda ^{4}m_{b} \\ 
\\ 
m_{s}=k_{s}\ \lambda ^{2}m_{b}%
\end{array}
\label{ks}
\end{equation}%
with phenomenologically, $k_{d}$ and $k_{s}$ being factores of order one.
Writing $W^{\dagger }$ in Eq.~\eqref{diag} as $W^{\dagger
}=(v_{1},v_{2},v_{3})$, with the $v_{i}$ vectors formed by the $i$-th column
of $W^{\dagger }$, we find e.g. for pattern II, 
\begin{equation}
v_{3}=\frac{1}{n_{3}}\left( 
\begin{array}{r}
\frac{m_{d}}{m_{b}}V_{11} \\ 
\frac{m_{s}}{m_{b}}V_{12} \\ 
V_{13}%
\end{array}%
\right) \times \left( 
\begin{array}{r}
\frac{m_{d}}{m_{b}}V_{31} \\ 
\frac{m_{s}}{m_{b}}V_{32} \\ 
V_{33}%
\end{array}%
\right)  \label{v3}
\end{equation}%
where $n_{3}$ is the norm of the vector obtained from the external product
of the two vectors. Taking into account the extra freedom of transformation
of the right-handed fields, we may choose $M_{31}^{d}=0$, corresponding to $%
c_{1}=0$ in Table~\ref{tab:downTextures}, and we conclude that%
\begin{equation}
v_{1}=\frac{1}{n_{1}}\left( 
\begin{array}{r}
\frac{m_{d}}{m_{b}}V_{31} \\ 
\frac{m_{s}}{m_{b}}V_{32} \\ 
V_{33}%
\end{array}%
\right) \times v_{3}^{\ast }  \label{v1}
\end{equation}%
Obviously, then $v_{2}=\frac{1}{n_{2}}v_{1}^{\ast }\times v_{3}^{\ast }$.
This process is replicated for all patterns. Thus, $V$ and $W$, are entirely
expressed in terms of Wolfenstein parameters and $k_{d}$ and $k_{s}$ of Eq.~%
\eqref{ks}. These two matrices will be later used to compute the patterns of
the FCNC's in Table \ref{tb:FCNCpatterns}. Indeed, in this way, we find e.g.
for pattern II, \ in leading order order, 
\begin{equation}
M_{d}=m_{b}\ \left( 
\begin{array}{ccc}
-k_{d}\ \lambda ^{3} & \left( \overline{\rho }-i\,\overline{\eta }\right) \
A\ \lambda ^{3} & 0 \\ 
-k_{d}\ \lambda ^{2} & A\ \lambda ^{2} & -k_{s}\ \lambda ^{3} \\ 
0 & 1 & 0%
\end{array}%
\right)  \label{mdl}
\end{equation}%
which corresponds to the expected power series where the couplings in $%
\Gamma _{1}$ to the first Higgs $\phi _{1}$ are comparatively smaller than
then couplings in $\Gamma _{2}$, and these smaller to\ the couplings in $%
\Gamma _{3}$. Similar results are obtained for all patterns in Table~\ref%
{tab:downTextures}, except for patterns IV, VII and X, where e.g. for
pattern IV, we find that the coupling in $(\Gamma _{1})_{33}$ is
proportional to$\ \lambda $, which is too large and contradicts our initial
assumption that all couplings in $\Gamma _{1}$ to the first Higgs $\phi _{1}$
must be smaller than the couplings in $\Gamma _{2}$ to the second Higgs $%
\phi _{2}$. Therefore, we exclude Patterns IV, VII and X.

We give in Table {\ref{Yukawa_example}} a numerical example of a Yukawa
coupling configuration for each pattern. We use the following quark running
masses at the electroweak scale $M_{Z}$: 
\begin{subequations}
\begin{align}
m_{u}& =1.3_{-0.2}^{+0.4}\,\text{MeV},\quad m_{d}=2.7\pm 0.3\,\text{MeV}%
,\quad m_{s}=55_{-3}^{+5}\,\text{MeV}, \\
m_{c}& =0.63\pm 0.03\,\text{GeV},\quad m_{b}=2.86_{-0.04}^{+0.05}\,\text{GeV}%
,\quad m_{t}=172.6\pm 1.5\,\text{GeV}.
\end{align}%
which were obtained from a renormalisation group equation evolution at
four-loop level \cite{1674-1137-38-9-090001}, which, taking into account all
experimental constrains \cite{Charles:2015gya}, implies: 
\end{subequations}
\begin{subequations}
\begin{align}
\lambda & =0.2255\pm 0.0006,\qquad A=0.818\pm 0.015, \\
\overline{\rho }& =0.124\pm 0.024,\qquad \overline{\eta }=0.354\pm 0.015.
\end{align}

\begin{table}[]
\caption{A numerical example of a Yukawa coupling configuration for each
pattern that gives the correct hierarchy among the quark masses and mixing.}{%
} {\label{Yukawa_example}} 
\par
\begin{center}
\setlength{\tabcolsep}{0.5pc} 
\resizebox{\textwidth}{!}{\begin{tabular}{|c|c|c|c|c|c|}
\hline
Pattern  & $v_1Y_1$ & $v_2Y_2$ & $v_3Y_3$ &  $M_d$ \\
\hline
I & 
$\begin{pmatrix}
	0.00277 & 0.0124 & 0.0101\,e^{1.907 \,i} \\
	0 & 0 & 0 \\
	0 & 0 & 0
\end{pmatrix}$
 & 
$\begin{pmatrix}
	0 & 0 & 0 \\
	0 & 0.0537 & 0.119 \\
	0 & 0 & 0
\end{pmatrix}$
 & 
$\begin{pmatrix}
	0 & 0 & 0 \\
	0 & 0 & 0 \\
	0 & 0 & 2.86
\end{pmatrix}$
 & 
$\begin{pmatrix}
	0.00277 & 0.00124 & 0.0101\,e^{1.907 \,i} \\
	0 & 0.0537 & 0.119 \\
	0 & 0 & 2.86
\end{pmatrix}$
\\
\hline
II  &
$\begin{pmatrix}
	0.0123 &  0.0101\,e^{-1.235 \,i} & 0 \\
	0 & 0 & 0.012 \\
	0 & 0 & 0
\end{pmatrix}$
 & 
$\begin{pmatrix}
	0 & 0 & 0 \\
	0.0524 & 0.119 & 0 \\
	0 & 0 & 0
\end{pmatrix}$
 & 
$\begin{pmatrix}
	0 & 0 & 0 \\
	0 & 0 & 0 \\
	0 & 2.86 & 0
\end{pmatrix}$
 & 
$\begin{pmatrix}
	0.0123 &  0.0101\,e^{-1.235 \,i} & 0 \\
	0.0524 & 0.119 & 0.012 \\
	0 & 2.86 & 0
\end{pmatrix}$
\\
\hline
III  &
$\begin{pmatrix}
	0.0127 & 0.0102\, e^{-1.253 \,i}& 0 \\
	0 & 0 & 0 \\
	0 & 0 & 0
\end{pmatrix}$
 & 
$\begin{pmatrix}
	0 & 0 & 0 \\
	0.0523 & 0.120 & 0 \\
	0 & 0 & 0.295
\end{pmatrix}$
 & 
$\begin{pmatrix}
	0 & 0 & 0 \\
	0 & 0 & 0 \\
	0 & 2.844 & 0
\end{pmatrix}$
 & 
$\begin{pmatrix}
	0.0127 & 0.0102\, e^{-1.253 \,i}  & 0 \\
	0.0523 & 0.120  & 0 \\
	0 & 2.844 & 0.295
\end{pmatrix}$
\\
\hline
V  &
$\begin{pmatrix}
	0.0127 & 0 & 0.0101\,e^{-1.234 \,i}\\
	0 & 0.0117  & 0 \\
	0 & 0 & 0
\end{pmatrix}$
 & 
$\begin{pmatrix}
	0 & 0 & 0 \\
	0.0524 & 0 & 0.112 \\
	0 & 0 & 0
\end{pmatrix}$
 & 
$\begin{pmatrix}
	0 & 0 & 0 \\
	0 & 0 & 0 \\
	0 & 0 & 2.86
\end{pmatrix}$
 & 
$\begin{pmatrix}
	0.0127 & 0 &0.0101\,e^{-1.234 \,i} \\
	0.0524  & 0.0117& 0.112 \\
	0 & 0 & 2.86
\end{pmatrix}$
\\
\hline
VI &
$\begin{pmatrix}
	0.0127 & 0 & 0.0102\,e^{-1.253 \,i} \\
	0 & 0 & 0 \\
	0 & 0 & 0
\end{pmatrix}$
 & 
$\begin{pmatrix}
	0 & 0 & 0 \\
	0.0523 & 0 & 0.120 \\
	0 & 0.295  & 0
\end{pmatrix}$
 & 
$\begin{pmatrix}
	0 & 0 & 0 \\
	0 & 0 & 0 \\
	0 & 0 & 2.844
\end{pmatrix}$
 & 
$\begin{pmatrix}
	0.0127 & 0 & 0.0102\,e^{-1.253 \,i} \\
0.0523 & 0 & 0.120 \\
	0 & 0.295 & 2.844
\end{pmatrix}$
\\
\hline
VIII &
$\begin{pmatrix}
	0 &0.0127 & 0.0102\,e^{1.907 \,i} \\
	0.0117 & 0 & 0 \\
	0 & 0 & 0
\end{pmatrix}$
 & 
$\begin{pmatrix}
	0 & 0 & 0 \\
	0 & 0.0524 & 0.119 \\
	0 & 0 & 0
\end{pmatrix}$
 & 
$\begin{pmatrix}
	0 & 0 & 0 \\
	0 & 0 & 0 \\
	0 & 0 & 2.86
\end{pmatrix}$
 & 
$\begin{pmatrix}
	0 &0.0127 & 0.0102\,e^{1.907 \,i} \\
	0.0117 & 0.0524 & 0.119 \\
	0 & 0 & 2.86
\end{pmatrix}$
\\
\hline
IX &
$\begin{pmatrix}
	0 & 0.0127 & 0.0101\,e^{-1.253 \,i} \\
	0 & 0 & 0 \\
	0 & 0 & 0
\end{pmatrix}$
 & 
$\begin{pmatrix}
	0 & 0 & 0 \\
	0 & 0.0523& 0.120 \\
	0.295  & 0 & 0
\end{pmatrix}$
 & 
$\begin{pmatrix}
	0 & 0 & 0 \\
	0 & 0 & 0 \\
	0 & 0 &2.844
\end{pmatrix}$
 & 
$\begin{pmatrix}
	0 & 0.0127 & 0.0101\,e^{-1.253 \,i} \\
	0 & 0.0523& 0.120 \\
	0.295& 0 &2.844
\end{pmatrix}$
\\
\hline
\end{tabular}}
\end{center}
\end{table}

\section{Predictions of flavour changing neutral currents}\label{sec:fcnc}

In the SM, flavour changing neutral currents (FCNC) are forbidden at tree
level, both in the gauge and the Higgs sectors. However, by extending the SM
field content, one obtains Higgs Flavour Violating Neutral Couplings \cite%
{Branco:2011iw}. In terms of the quark mass eigenstates, the Yukawa couplings
to the Higgs neutral fields are:

\end{subequations}
\begin{equation}
\begin{aligned} -\mathcal{L}_{\text{Neutral Yukawa}}= &\frac{H_0}{v}\left(
\overline{d_L}\, D_d \, d_R + \overline{u_L}\, D_u \, u_R \right) +
\frac{1}{v'} \overline{d_L} \, N^d_{1}\, \left( R_1 + i\, I_1 \right) \, d_R
\\ & + \frac{1}{v'} \overline{u_L} \, N^u_{1} \, \left( R_1 - i\, I_1
\right) \, u_R + \frac{1}{v''} \overline{d_L} \, N^d_{2}\, \left( R_2 + i\,
I_2 \right) \, d_R \\ &+ \frac{1}{v''} \overline{u_L} \, N^u_{2} \, \left(
R_2 - i\, I_2 \right) \, u_R + h.c. \end{aligned}
\end{equation}%
where the $N_{i}^{u,d}$ are the matrices which give the strength and the
flavour structure of the FCNC, 
\begin{subequations}
\begin{align}  \label{eq:FCNC}
& N_{1}^{d}=\frac{1}{\sqrt{2}}V^{\dagger }\,\left( v_{2}\Gamma
_{1}-v_{1}e^{i\,\alpha _{2}}\Gamma _{2}\right) \,W, \\
& N_{2}^{d}=\frac{1}{\sqrt{2}}V^{\dagger }\left( v_{1}\Gamma
_{1}+v_{2}e^{i\,\alpha _{2}}\Gamma _{2}-\frac{v_{1}^{2}+v_{2}^{2}}{v_{3}}%
e^{i\,\alpha _{3}}\Gamma _{3}\right) \,W, \\
& N_{1}^{u}=\frac{1}{\sqrt{2}}\left( v_{2}\Omega _{1}-v_{1}e^{-i\,\alpha
_{2}}\Omega _{2}\right) , \\
& N_{2}^{u}=\frac{1}{\sqrt{2}}\left( v_{1}\Omega _{1}+v_{2}e^{-i\,\alpha
_{2}}\Omega _{2}-\frac{v_{1}^{2}+v_{2}^{2}}{v_{3}}e^{-i\,\alpha _{3}}\Omega
_{3}\right) .
\end{align}%
Since in our case the $N_{i}^{u}$ are diagonal, there are no flavour
violating terms in the up-sector. Therefore, the analysis of the FCNC
resumes only to the down-quark sector. One can use the equations of the mass
matrices presented in Eq.~\eqref{eq:mass} to simplify the Higgs mediated
FCNC matrices for the down-sector: 
\end{subequations}
\begin{subequations}
\label{eq:simplefcnc}
\begin{align}
N_{1}^{d}& =\frac{v_{2}}{v_{1}}D_{d}-\frac{v_{2}}{\sqrt{2}}\left( \frac{v_{2}%
}{v_{1}}+\frac{v_{1}}{v_{2}}\right) e^{i\alpha _{2}}\,V^{\dagger }\,\Gamma
_{2}\,W-\frac{v_{2}\,v_{3}}{v_{1}\sqrt{2}}e^{i\alpha _{3}}V^{\dagger
}\,\,\Gamma _{3}\,W \\[2mm]
N_{2}^{d}& =D_{d}-\frac{v^{2}}{v_{3}\sqrt{2}}e^{i\alpha _{3}}\,V^{\dagger
}\,\Gamma _{3}\,W
\end{align}

In order to satisfy experimental constraints arising from $K^{0}-\overline{%
K^{0}}$, $B^{0}-\overline{B^{0}}$ and $D^{0}-\overline{D^{0}}$, the
off-diagonal elements of the Yukawa interactions $N_{1}^{d}$ and $N_{2}^{d}$
must be highly suppressed \cite{Botella:2014ska} \cite{AndreasCrivellin2013}%
. For each of our 10 solutions in Table~\ref{tab:downTextures}, we summarize
in Table {\ref{tb:FCNCpatterns}} all FCNC patterns, for each solution, and
for $v_{1}=v_{2}=v_{3}$ and $\alpha _{2}=\alpha _{3}=0$. These patterns are
of the BGL type, since in Eq.~\eqref{eq:simplefcnc} all matrices can be
expressed in terms of the CKM mixing matrix elements and the down quark
masses. As explained, to obtain these patterns, we express the CKM matrix $V$
and the matrix $W$ in terms of Wolfenstein parameters.

\begin{table}[]
\caption{For all allowed patterns, we find that the matrices $N^d_1-D_d$ and 
$N^d_2$ are proportional to the following paterns, where $\protect\lambda$
is the Cabibbo angle.}{\label{tb:FCNCpatterns}} 
{\ } \setlength{\tabcolsep}{14pt} \centering 
\begin{tabular}{|c|c|c|}
\hline
Pattern & $(N^d_1-D_d)\sim$ & $N^d_{2}\sim$ \\ \hline
I & $%
\begin{pmatrix}
\lambda^4 & \lambda^3 & \lambda^3 \\ 
\lambda^5 & \lambda^2 & \lambda^2 \\ 
\lambda^7 & \lambda^4 & 1%
\end{pmatrix}%
$ & $%
\begin{pmatrix}
\lambda^4 & \lambda^7 & \lambda^3 \\ 
\lambda^9 & \lambda^2 & \lambda^2 \\ 
\lambda^7 & \lambda^4 & 1%
\end{pmatrix}
$ \rule{0pt}{0.9cm}\rule[-0.9cm]{0pt}{0pt} \\ 
II & $%
\begin{pmatrix}
\lambda^4 & \lambda^3 & \lambda^3 \\ 
\lambda^3 & \lambda^2 & \lambda^2 \\ 
\lambda^5 & \lambda^4 & 1%
\end{pmatrix}%
$ & $%
\begin{pmatrix}
\lambda^4 & \lambda^7 & \lambda^3 \\ 
\lambda^9 & \lambda^2 & \lambda^2 \\ 
\lambda^7 & \lambda^4 & 1%
\end{pmatrix}%
$ \rule{0pt}{0.9cm}\rule[-0.9cm]{0pt}{0pt} \\ 
III & $%
\begin{pmatrix}
\lambda^4 & \lambda^3 & \lambda^3 \\ 
\lambda^3 & \lambda^2 & \lambda^2 \\ 
\lambda & \lambda^2 & 1%
\end{pmatrix}%
$ & $%
\begin{pmatrix}
\lambda^4 & \lambda^5 & \lambda^3 \\ 
\lambda^3 & \lambda^2 & \lambda^2 \\ 
\lambda & \lambda^2 & 1%
\end{pmatrix}%
$ \rule{0pt}{0.9cm}\rule[-0.9cm]{0pt}{0pt} \\ 
IV & $%
\begin{pmatrix}
\lambda^4 & \lambda^3 & \lambda^3 \\ 
\lambda^3 & \lambda^2 & \lambda^2 \\ 
\lambda & \lambda^2 & 1%
\end{pmatrix}%
$ & $%
\begin{pmatrix}
\lambda^4 & \lambda^5 & \lambda^3 \\ 
\lambda^3 & \lambda^2 & \lambda^2 \\ 
\lambda & \lambda^2 & 1%
\end{pmatrix}%
$\rule{0pt}{0.9cm}\rule[-0.9cm]{0pt}{0pt} \\ 
V & $%
\begin{pmatrix}
\lambda^4 & \lambda^3 & \lambda^3 \\ 
\lambda^3 & \lambda^2 & \lambda^2 \\ 
\lambda^5 & \lambda^4 & 1%
\end{pmatrix}%
$ & $%
\begin{pmatrix}
\lambda^4 & \lambda^7 & \lambda^3 \\ 
\lambda^7 & \lambda^2 & \lambda^2 \\ 
\lambda^5 & \lambda^4 & 1%
\end{pmatrix}%
$ \rule{0pt}{0.9cm}\rule[-0.9cm]{0pt}{0pt} \\ 
VI & $%
\begin{pmatrix}
\lambda^4 & \lambda^3 & \lambda^3 \\ 
\lambda^3 & \lambda^2 & \lambda^2 \\ 
\lambda & \lambda^2 & 1%
\end{pmatrix}%
$ & $%
\begin{pmatrix}
\lambda^4 & \lambda^5 & \lambda^3 \\ 
\lambda^3 & \lambda^2 & \lambda^2 \\ 
\lambda & \lambda^2 & 1%
\end{pmatrix}%
$ \rule{0pt}{0.9cm}\rule[-0.9cm]{0pt}{0pt} \\ 
VII & $%
\begin{pmatrix}
\lambda^4 & \lambda^3 & \lambda^3 \\ 
\lambda^3 & \lambda^2 & \lambda^2 \\ 
\lambda & \lambda^2 & 1%
\end{pmatrix}%
$ & $%
\begin{pmatrix}
\lambda^4 & \lambda^5 & \lambda^3 \\ 
\lambda^3 & \lambda^2 & \lambda^2 \\ 
\lambda & \lambda^2 & 1%
\end{pmatrix}%
$ \rule{0pt}{0.9cm}\rule[-0.9cm]{0pt}{0pt} \\ 
VIII & $%
\begin{pmatrix}
\lambda^4 & \lambda^3 & \lambda^3 \\ 
\lambda^3 & \lambda^2 & \lambda^2 \\ 
\lambda^5 & \lambda^4 & 1%
\end{pmatrix}%
$ & $%
\begin{pmatrix}
\lambda^4 & \lambda^7 & \lambda^3 \\ 
\lambda^7 & \lambda^2 & \lambda^2 \\ 
\lambda^5 & \lambda^4 & 1%
\end{pmatrix}%
$ \rule{0pt}{0.9cm}\rule[-0.9cm]{0pt}{0pt} \\ 
IX & $%
\begin{pmatrix}
\lambda^4 & \lambda^3 & \lambda^3 \\ 
\lambda^3 & \lambda^2 & \lambda^2 \\ 
\lambda & \lambda^2 & 1%
\end{pmatrix}%
$ & $%
\begin{pmatrix}
\lambda^4 & \lambda^5 & \lambda^3 \\ 
\lambda^3 & \lambda^2 & \lambda^2 \\ 
\lambda & \lambda^2 & 1%
\end{pmatrix}%
$ \rule{0pt}{0.9cm}\rule[-0.9cm]{0pt}{0pt} \\ 
X & $%
\begin{pmatrix}
\lambda^4 & \lambda^3 & \lambda^3 \\ 
\lambda^3 & \lambda^2 & \lambda^2 \\ 
\lambda & \lambda^2 & 1%
\end{pmatrix}%
$ & $%
\begin{pmatrix}
\lambda^4 & \lambda^5 & \lambda^3 \\ 
\lambda^3 & \lambda^2 & \lambda^2 \\ 
\lambda & \lambda^2 & 1%
\end{pmatrix}%
$\rule{0pt}{0.9cm}\rule[-0.9cm]{0pt}{0pt} \\ \hline
\end{tabular}%
\end{table}

The tree level Higgs mediated $\Delta S=2$ amplitude must be suppressed.
This may allways be achieved if one chooses the masses of the flavour
violating neutral Higgs scalars sufficiently heavy. However, from the
experimental point of view, it would be interesting to have these masses as
low as possible. Therefore, we also estimate the lower bound of these
masses, by considering the contribution to $B^{0}-\overline{B^{0}}$ mixing.
We choose this mixing, since for our patterns, the $(3,1)$ entry of the
matrix $N_{1}^{d}$ is the less suppressed in certain cases and would require
very heavy flavour violating neutral Higgses. The relevant quantity is the
off-diagonal matrix element $M_{12}$, which connects the B meson with the
corresponding antimeson. This matrix element, $M_{12}^{NP}$, receives
contributions \cite{Botella:2014ska} both from a SM box diagram and a
tree-level diagram involving the FCNC: 
\end{subequations}
\begin{equation}
M_{12}=M_{12}^{SM}+M_{12}^{NP},
\end{equation}%
where the New Physics (NP) short distance tree level contribution to the meson-antimeson
contribution is: 
\begin{equation}
\begin{aligned} M_{12}^{NP}=& \sum_i^2 {\frac{f_B^2 \, m_B}{ 96\, v^2
m^2_{R_i}} } \left\{ \left[ \left(1+ \left( \frac{m_B}{m_d+m_b} \right)^2
\right)\left(a^R_i\right)_{12} \right] - \left[ \left(1+ 11 \left(
\frac{m_B}{m_d+m_b} \right)^2 \right] \right)\left(b^R_i\right)_{12}
\right\} \\ &+\sum_i^2 {\frac{f_B^2 \, m_B}{ 96\, v^2 m^2_{I_i}} } \left\{
\left[ \left( 1+ \left( \frac{m_B}{m_d+m_b}
\right)^2\right)\left(a^I_i\right)_{12}\right] - \left[ \left(1+ 11 \left(
\frac{m_B}{m_d+m_b} \right)^2\right]\right) \left(b^I_i\right)_{12} \right\}
\end{aligned}
\end{equation}%
with $v^{2}=v_{1}^{2}+v_{2}^{2}+v_{2}^{2}$ and 
\begin{equation}
\begin{array}{l}
\left( a_{i}^{R}\right) _{12}=\left[ \left( N_{i}^{d}\right) _{31}^{\ast
}+\left( N_{i}^{d}\right) _{13}\right] ^{2} \\ 
\left( a_{i}^{I}\right) _{12}=-\left[ \left( N_{i}^{d}\right) _{31}^{\ast
}-\left( N_{i}^{d}\right) _{13}\right] ^{2}%
\end{array}%
~,\qquad 
\begin{array}{l}
\left( b_{i}^{R}\right) _{12}=\left[ \left( N_{i}^{d}\right) _{31}^{\ast
}-\left( N_{i}^{d}\right) _{13}\right] ^{2} \\ 
\left( b_{i}^{I}\right) _{12}=-\left[ \left( N_{i}^{d}\right) _{31}^{\ast
}+\left( N_{i}^{d}\right) _{13}\right] ^{2}%
\end{array}%
~,\qquad i=1,2  \label{ab}
\end{equation}%
In order to obtain a conservative measure, we have tentatively expanded the
original expression in \cite{Botella:2014ska} and, for the three Higgs case,
included all neutral Higgs mass eigenstates.

Adopting as input values the PDG experimental determinations of $f_{B}$, $%
m_{B}$ and $\Delta \,m_{B}$ and considering a common VEV for all Higgs
doublets, we impose the inequality $M_{12}^{NP}<\Delta m_{B}$. The following
plots show an estimate of the lower bound for the flavour-violating Higgs
masses for two different patterns. We plot two masses chosen from the set $%
\left( m_{1}^{R},m_{2}^{R},m_{1}^{I},m_{2}^{I}\right) $, while the other two
are varied over a wide range. In Fig. 1, we illustrate these lower bounds
for Pattern III, which are restricted by the $(3,1)$ entry of $N_{1}^{d}$
matrix and suppressed by a factor of $\lambda $. For Pattern VIII, in Fig. 2
we find the flavour violating neutral Higgs to be much lighter and possibly
accessible at LHC.

\section{Conclusions}

\label{sec:conc}

We have presented a model based on the SM with 3 Higgs and an additional
flavour discrete symmetry. We have shown that there exist flavour discrete
symmetry configurations which lead to the alignment of the quark sectors.\
By allowing each scalar field to couple to each quark generation with a
distinctive scale, one obtains the quark mass hierarchy, and although this
hierarchy does not arise from the symmetry, the effect of both is such that
the CKM matrix is near to the identity and has the correct overall
phenomenological features. In this context, we have obtained 7 solutions
fulfilling these requirements, with the additional constraint of the up
quark mass matrix being diagonal and real.

We have also verified if accidental $U(1)$ symmetries may appear in the
Yukawa sector or in the potential, particularly the case where a continuous
accidental $U(1)$ symmetry could arise, once the $Z_{7}$ is imposed at the
Lagrangian level. This was indeed the case, however we shown that the
anomaly-free conditions of global symmetries are violated. Thus, the global $%
U(1)_{X}$ symmetry is anomalous and therefore only the discrete symmetry $%
Z_{7}$ persists.

As in this model new Higgs doublets are added, one expects large FCNC
effects, already present at tree level. However, such effects have not been
experimentally observed. We show that, for certain of our specific
implementations of the flavour symmetry, it is possible to suppress the FCNC
effects and to ensure that the flavour violating neutral Higgs are
light enough to be detectable at LHC. Indeed, in this respect, our model
is a generalization of the BGL models for 3HDM, since the FCNC flavour
structure is entirely determined by CKM.

\begin{figure}[h!]
\centering
\begin{subfigure}{.5\textwidth}
  \centering
  \includegraphics[width=1.\linewidth]{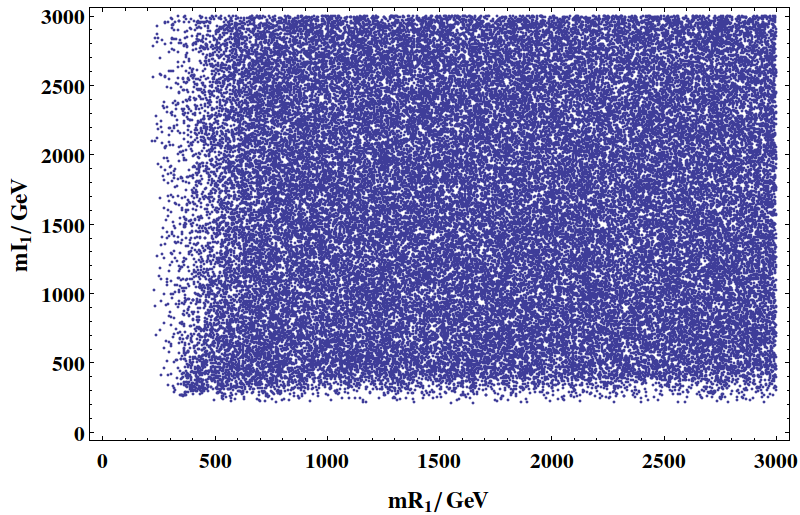}
  \caption{Estimate of the lower bound for the flavour-violating Higgs masses for $R_1$ and $I_1$.}
  \label{fig:sub1}
\end{subfigure}%
\begin{subfigure}{.5\textwidth}
  \centering
  \includegraphics[width=1.\linewidth]{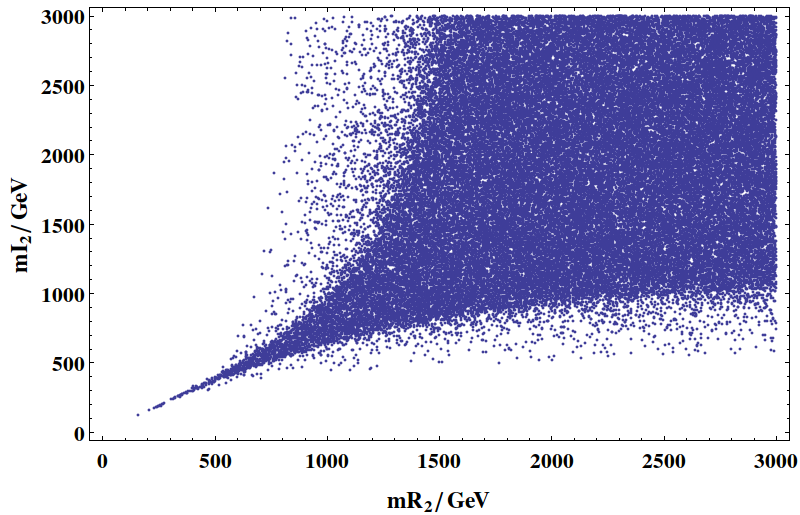}
  \caption{Estimate of the lower bound for the flavour-violating Higgs masses for $R_2$ and $I_2$.}
  \label{fig:sub2}
\end{subfigure}
\caption{Lower bound for the flavour-violating Higgs masses for case III. }
\label{fig:test}
\end{figure}

\begin{figure}[h!]
\centering
\begin{subfigure}{.5\textwidth}
  \centering
  \includegraphics[width=1.\linewidth]{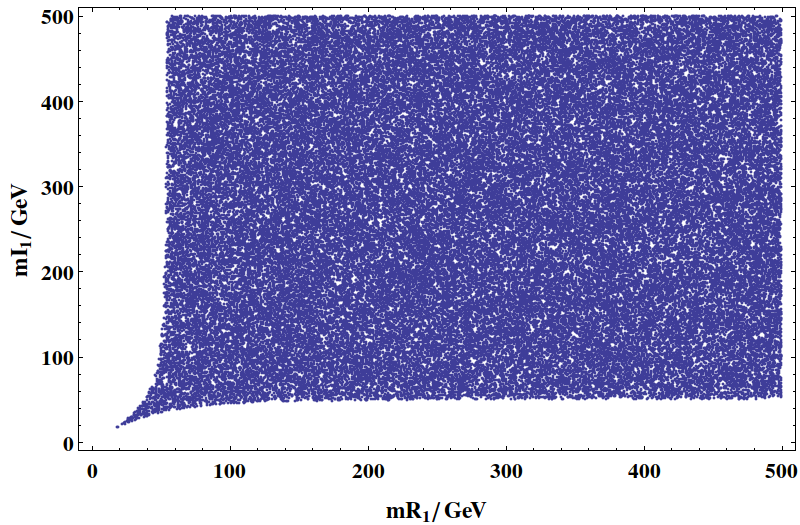}
  \caption{Estimate of the lower bound for the flavour-violating Higgs masses for $R_1$ and $I_1$.}
  \label{fig:sub1}
\end{subfigure}%
\begin{subfigure}{.5\textwidth}
  \centering
  \includegraphics[width=1.\linewidth]{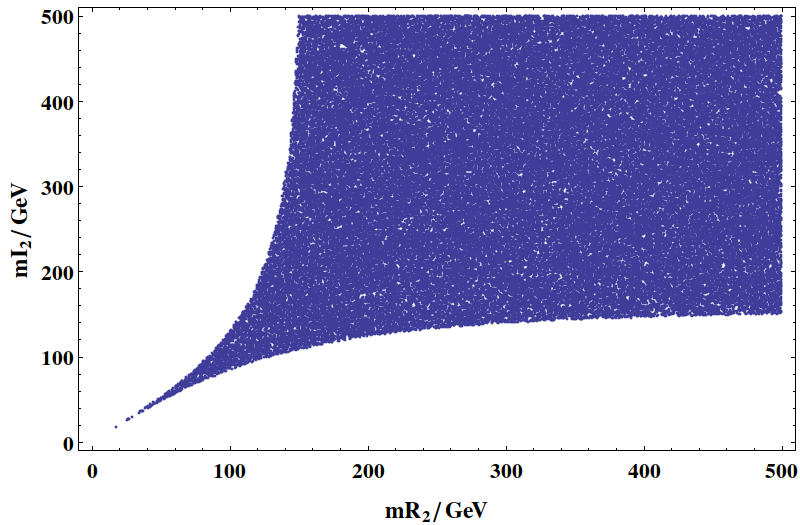}
  \caption{Estimate of the lower bound for the flavour-violating Higgs masses for $R_2$ and $I_2$.}
  \label{fig:sub2}
\end{subfigure}
\caption{Lower bound for the flavour-violating Higgs masses for case VIII. }
\label{fig:test1}
\end{figure}
%%%%%%%%%%%%%%%%%%%%%%%%%%%%%%%%%%%%%%%%%%%%%%%%%%%%%%%%%%%%%%%%%%%%%%
\acknowledgments

This work is partially supported by Funda\c{c}\~{a}o para a Ci\^{e}ncia e a
Tecnologia (FCT, Portugal) through the projects CERN/FP/123580/2011,
PTDC/FIS-NUC/0548/2012, EXPL/FIS-NUC/0460/2013, and CFTP-FCT Unit 777
(PEst-OE/FIS/UI0777/2013) which are partially funded through POCTI (FEDER),
COMPETE, QREN and EU. The work of D.E.C. is also supported by Associa\c c\~
ao do Instituto Superior T\'ecnico para a Investiga\c c\~ao e
Desenvolvimento (IST-ID). N.R.A is supported by European Union Horizon 2020
research and innovation programme under the Marie Sklodowska-Curie grant
agreement No 674896. N.R.A is grateful to CFTP for the hospitality during
his stay in Lisbon.

\bibliographystyle{ieeetr}
\bibliography{refs}

\end{document}